\documentclass[aps, prb, reprint, showpacs]{revtex4-1}
\usepackage{amsmath}
\usepackage[pdftex]{graphicx}
\usepackage{graphicx}
\usepackage{amsmath}
\usepackage{amssymb}
\usepackage{url}
\usepackage{bm}		
\usepackage{color}					

\begin{document}

%\preprint{}

\title{A compact, integrated silicon device for the generation of spectrally-filtered, pair-correlated photons} %Title of paper



\author{Momchil Minkov}
\email[]{momchil.minkov@epfl.ch}

\author{Vincenzo Savona}
\affiliation{Laboratory of Theoretical Physics of Nanosystems, Ecole Polytechnique F\'{e}d\'{e}rale de Lausanne (EPFL), CH-1015 Lausanne, Switzerland}

\date{\today}

\begin{abstract}
The third-order non-linearity of silicon gives rise to a spontaneous four-wave mixing process in which correlated photon pairs are generated. Sources based on this effect can be used for quantum computation and cryptography, and can in principle be integrated with standard CMOS fabrication technology and components. However, one of the major challenges is the on-chip demultiplexing of the photons, and in particular the filtering of the pump power, which is many orders of magnitude larger than that of the signal and idler photons. Here, we propose a photonic crystal coupled-cavity system designed so that the coupling of the pump mode to the output channel is strictly zero due to symmetry. We further analyze this effect in presence of fabrication disorder and find that, even then, a pump suppression of close to 40 dB can be achieved in state-of-the art systems. Due to the small mode volumes and high quality factors, our system is also expected to have a generation efficiency much higher than in standard micro-ring systems. Those two considerations make a strong case for the integration of our proposed design in future on-chip quantum technologies. 
\end{abstract}

% \pacs{73.43.-f, 42.70.Qs, 03.65.Vf}% insert suggested PACS numbers in braces on next line

\maketitle %\maketitle must follow title, authors, abstract and \pacs

% Introduction
\section{Introduction}

Photons are expected to play a major role in future quantum computation technologies, serving as qubits for processing and/or for transferring of quantum information \cite{Knill2001, Kimble2008, OBrien2009}. One of the mandatory requirements for these operations is a source of indistinguishable single photons. Furthermore, a source of correlated photon pairs would bring an additional advantage to the functionalities of such a photonic platform. Ideally, the platform should also be compatible with current Complementary Metal-oxide-semiconductor (CMOS) production techniques to ensure cost-effectiveness, scalability, and integrability with standard electronic components \cite{Sun2015}. Because of that, correlated photon-pair sources based on the third-order non-linearity of silicon have been widely studied \cite{Lin2006, Sharping2006, Clemmen2009, Helt2010, Matsuda2011, Xiong2011, Azzini2012, Azzini2012a, Azzini2013, Harris2014, Gentry2015}, and their potential to serve as \textit{heralded} single-photons sources \cite{Davancco2012, Collins2013}, as well as sources of \textit{entangled} photon states \cite{Takesue2007, Chen2011, Takesue2014, Grassani2015} has been demonstrated.  

Correlated photon pairs in silicon are created through spontaneous four-wave mixing (FWM) whereby two photons from a pump source are converted into a signal and idler photon under appropriate phase-matching conditions. This process can also be stimulated, in which case input power at the signal frequency increases the output power of the idler frequency, which can be used for parametric oscillation \cite{Lin2007, Monat2010, Shinkawa2011, Zeng2014}. However, it is only through the spontaneous process that the correlated, indistinguishable photon pairs needed for quantum applications can be generated. The efficiency of the FWM can be largely improved through the use of nano-engineered materials, such as micro-ring resonators \cite{Clemmen2009, Chen2011, Azzini2012a, Davancco2012, Grassani2015}, or Photonic Crystal (PhC) slow-light waveguides \cite{Monat2010, Shinkawa2011, Xiong2011, Matsuda2011, Takesue2014} or cavities \cite{Azzini2013}. The PhC cavity platform is in fact largely unexplored, yet very promising, since ultra-high quality factors ($Q$) above one million can be achieved in ultra-small mode-volume (smaller than $(\lambda/n)^3$) cavities \cite{Minkov2014, Lai2014}. In addition, the compatibility of PhC devices with CMOS technologies has already been demonstrated \cite{Shinkawa2011, Mehta2014, Ooka2015}. 

\begin{figure}
\centering
\includegraphics[width = 0.46\textwidth, trim = 0in 0in 0in 0in, clip = true]{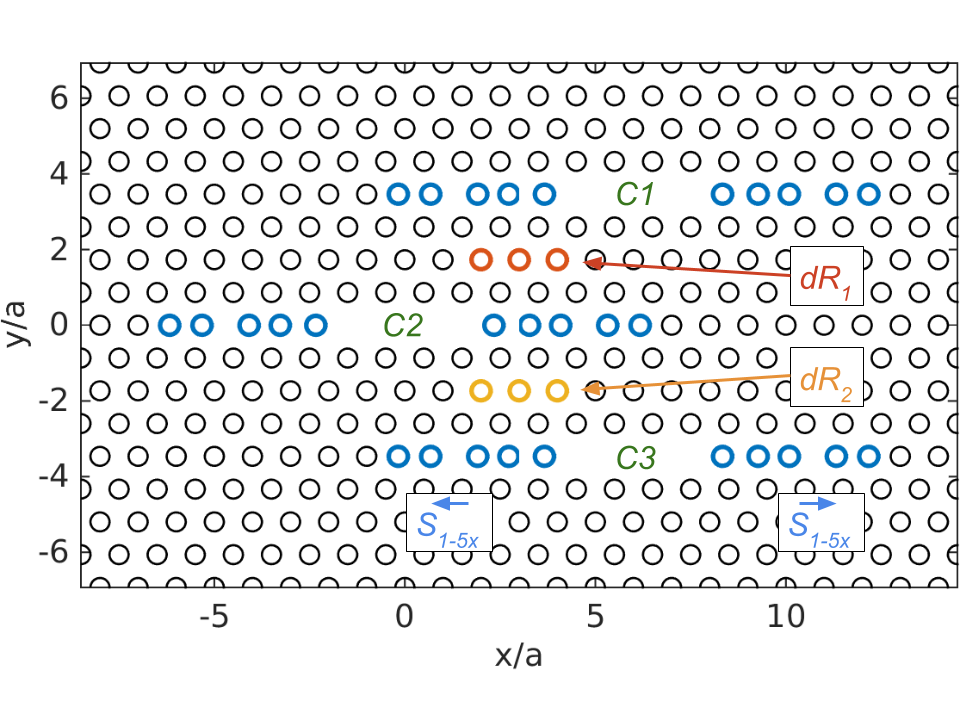}%
 \caption{The proposed PhC device for resonant FWM. There are three L3 cavities (marked $C1$, $C2$, $C3$) optimized for a high quality factor through the hole shifts $S_{1-5x}$ (marked in blue). Furthermore, we explore the possibility to tune the device through the radii of the holes marked in red (orange), which are changed by a value $dR_1$ ($dR_2$) from the starting PhC value $R$.}
\label{fig1}
\end{figure}

\begin{figure*}
\centering
\includegraphics[width = \textwidth, trim = 0in 0in 0in 0in, clip = true]{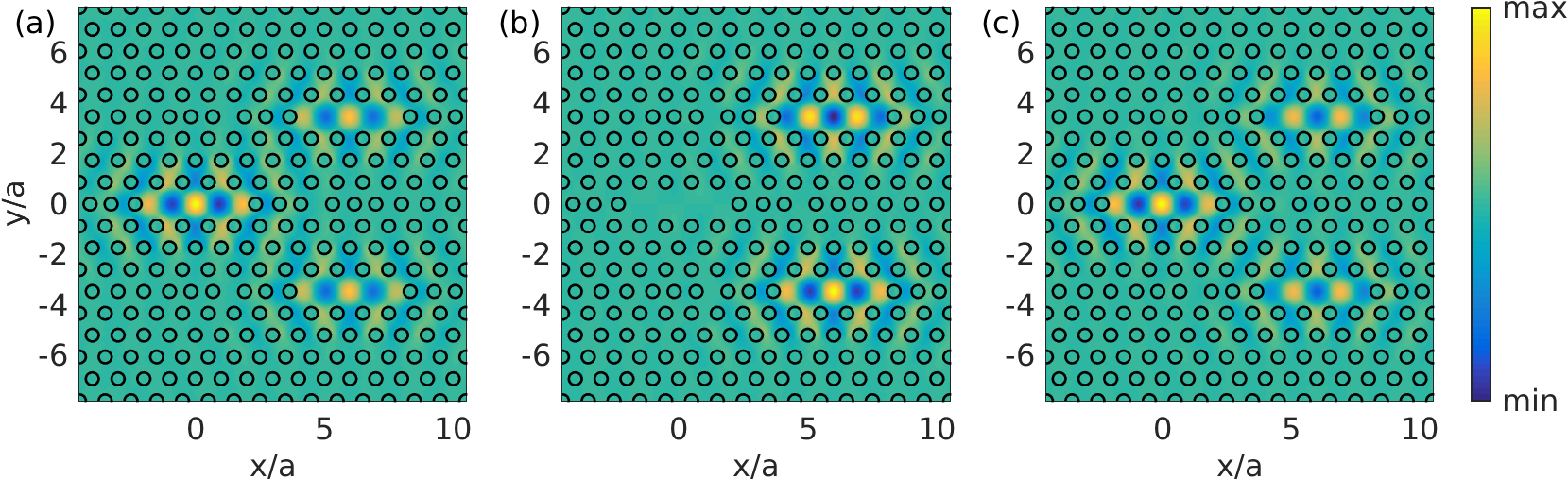}%
 \caption{Electric field $E_y$ component of the three eigenmodes of the device, with increasing frequency. (a): $M1$, at $\omega_s = 1.6154 a/c$; (b): $M2$, at $\omega_p = 1.6191 a/c$; (a): $M3$, at $\omega_i = 1.6228 a/c$. The shifts $S_{1-5x}$ are as given in Ref. [\onlinecite{Minkov2014}], while $dR_1 = dR_2 = 0$ (see also Fig. \ref{fig1}). The symmetry of the spatial profile of $E_y$ is opposite to the eigenvalue of the $\hat{\sigma}_{xz}$ reflection operator (see text). Thus, globally, M1 and M3 are antisymmetric, while M2 is symmetric.}
\label{fig2}
\end{figure*}

One of the major problems with respect to integrating FWM photon sources on-chip is the filtering of the pump power, which is typically orders of magnitude larger than the generated signal/idler power. Thus, the filtering has commonly been performed by an external band-pass filter. Very recently, the first device that achieves sufficiently strong on-chip filtering has been demonstrated \cite{Harris2014}, which, together with recent results on multiplexing \cite{Collins2013} and interfering \cite{Silverstone2013} silicon FWM-based sources, paves the way to large-scale applications. In Ref. [\onlinecite{Harris2014}], the on-chip filtering was achieved through an integrated Distributed Bragg Reflector (DBR) with a band-gap at the pump frequency. While the method proved effective, this DBR had the largest size among all individual components, thus significantly increasing the footprint of the setup.

Here, we propose a silicon photonic crystal device consisting of three coupled high-$Q$ cavities. The system supports three modes that define the signal, pump, and idler frequencies for a resonant-FWM process. We show that the device can be designed so that only the signal and idler modes couple to an output waveguide, while the pump mode has \textit{strictly zero} coupling due to a spatial symmetry. Coupling \textit{into} the pump mode can still be easily achieved through a waveguide that is off the symmetry axis. The self-filtering of the pump and the expected high efficiency of the non-linear process as compared to micro-ring resonators are both very promising features of our proposal. To analyze its qualities in practice, we also study the system in the presence of fabrication imperfections. These break the underlying spatial symmetry, and introduce a finite coupling of the pump mode to the output channel. However, we find that, for state-of-the-art disorder magnitudes, a difference of the transmission between pump and signal/idler of close to four orders of magnitude can still be expected. 

\section{Device}

The proposed design is based on a PhC made of a triangular lattice of circular holes of radius $R$ in a silicon slab of thickness $d$ suspended in air (Fig. \ref{fig1}). The values are set to $R = 0.25a$, $d = 0.55a$, with $a$ the lattice constant. The latter is kept as a free parameter, but the design is such that for the standard thickness of $d = 220\mathrm{nm}$ of the silicon slab, the operational wavelength is in the telecommunication window around $1.55\mathrm{\mu m}$. For the resonant FWM, three identical L3-type cavities are introduced, marked as $C1, C2$ and $C3$ in Fig. \ref{fig1}. We use the optimized L3 designs from Ref. [\onlinecite{Minkov2014}], where the five holes on each side of the cavity (marked in blue in the Figure) are shifted away by five different values ($S_{1-5x}$), such that the $Q$ is maximum. The shifts are $S_{1-5x} = [0.337, 0.270, 0.088, 0.323, 0.173]a$, and a self-standing cavity has a quality factor of $Q = 4.1 \times 10^6$. In our case, the cavity modes are coupled, and mix to form three eigenmodes of the structure, which we label $M1, M2, M3$. We will also refer to those as the signal, pump, and idler modes, respectively, and so we denote their frequencies as $\omega_s, \omega_p$, and $\omega_i$. The spatial distribution of the $y$-component of the electric field of these modes is shown in Fig. \ref{fig2}. The simulation was done using the guided-mode expansion (GME) method \cite{Andreani2006}. 

The crucial observation that enables the pump filtering is based on the symmetry of the structure with respect to reflection in the $xz$-plane. As can be seen from Fig. \ref{fig2}, the three modes have eigenvalues $-1$, $1$, and $-1$ with respect to the $\hat{\sigma}_{xz}$ reflection operator. Notice that, in the Figure, the $E_y$ electric field component is illustrated, which flips under $\hat{\sigma}_{xz}$, and thus the spatial profiles have a symmetry opposite to the eigenvalues quoted above. Nevertheless, the correct classification is with respect to the \textit{global} symmetry, which is why throughout this paper we refer to $M2$ as the \textit{symmetric} cavity mode, and to $M1$ and $M3$ as the \textit{anti-symmetric} ones. The symmetry of the eigenmodes of a three-cavity system has already been proposed as a means to provide some filtering of the pump mode in FWM systems \cite{Azzini2013, Zeng2015}. However, only the fact that the symmetric mode has very low electric field intensity in the central cavity was exploited in those proposals (see Fig. \ref{fig2}(b)), and not the opportunity to fully cancel the field, offered by the symmetry. This is why the filtering was only partial. Here instead, we will show that \textit{full} filtering of this mode can be achieved when the output channel does not support modes of the corresponding symmetry.

\begin{figure}
\centering
\includegraphics[width = 0.46\textwidth, trim = 0in 0in 0in 0in, clip = true]{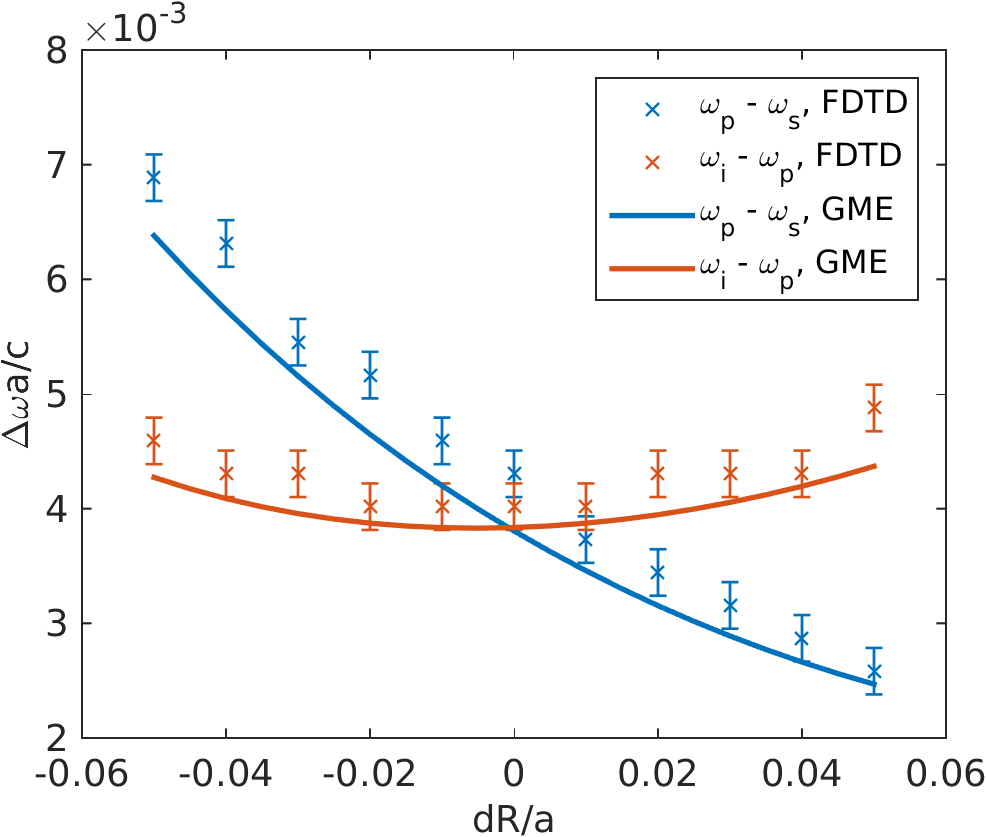}%
 \caption{Frequency difference between the first and the second modes ($\omega_p - \omega_s$) and between the second and the third modes ($\omega_i - \omega_p$) of the coupled-cavity system, as a function of a radius shift $dR_1 = dR_2 = dR$ (see Fig. \ref{fig1}).}
\label{fig3}
\end{figure}

In resonant FWM in a cavity system like ours, the phase-matching condition simply reduces to the energy-matching $2\omega_p = \omega_s + \omega_i$ (within a tolerance defined by the spectral linewidths). We must thus ensure that this condition can be met in our device. In order to allow for some tunability, we define two parameters, $dR_1$ and $dR_2$, corresponding to a radius shift of the three holes marked in red and orange, respectively, in Fig. \ref{fig1}. This is the simplest way to simulate a generic tuning of the intra-cavity coupling between the cavities $C_1$ and $C_2$ ($dR_1$), and $C_2$ and $C_3$ ($dR_2$). In practice this can also be achieved through a local modulation of the photonic environment, for example using electro-optic modulation \cite{Xu2005}, or selective surface oxidation \cite{Intonti2012}. The possibility for such a tuning is highly desirable since, even for perfectly-matched frequencies of the nominal structure, fabrication imperfections introduce random fluctuations of the cavity resonances \cite{Minkov2013}. In Section \ref{sec:disorder} we will in fact analyze the possibility to use these parameters to counter the negative effect of fabrication disorder, both in view of the frequency matching and in view of the pump filtering. As a first step here, we verify that the modes can be tuned to the necessary frequency relation already without disorder.

In Fig. \ref{fig3}, we plot the two frequency differences, $\omega_p - \omega_s$, and $\omega_i - \omega_p$, versus a shift $dR$ such that $dR_1 = dR_2 = dR$. At this stage we keep the two shifts equal in order to preserve the $xz$-symmetry. We will only study different shifts in Section \ref{sec:disorder}, where this symmetry is already broken by disorder. The frequencies were computed with the GME, as well as with a commercial-grade simulator based on the Finite-difference Time-domain (FDTD) method \cite{lumerical}. In the latter simulation, a broadband, $y$-polarized Gaussian source centered at $C_1$ was used to excite all three modes, and the spectrum was computed from the Fourier transform of the time-dependent electric field recorded in the same spot, after the decay of the source. The error bars are defined by the total simulation time, which was set to $25 \mathrm{ps}$. Very good agreement between the two computation methods is found, and, importantly, the two curves cross, demonstrating that the frequency matching can be fulfilled. In fact, the best nominal structure is the one with $dR_1 = dR_2 = 0$, i.e. with no extra hole modulation. 

\begin{figure}
\centering
\includegraphics[width = 0.46\textwidth, trim = 0in 0in 0in 0in, clip = true]{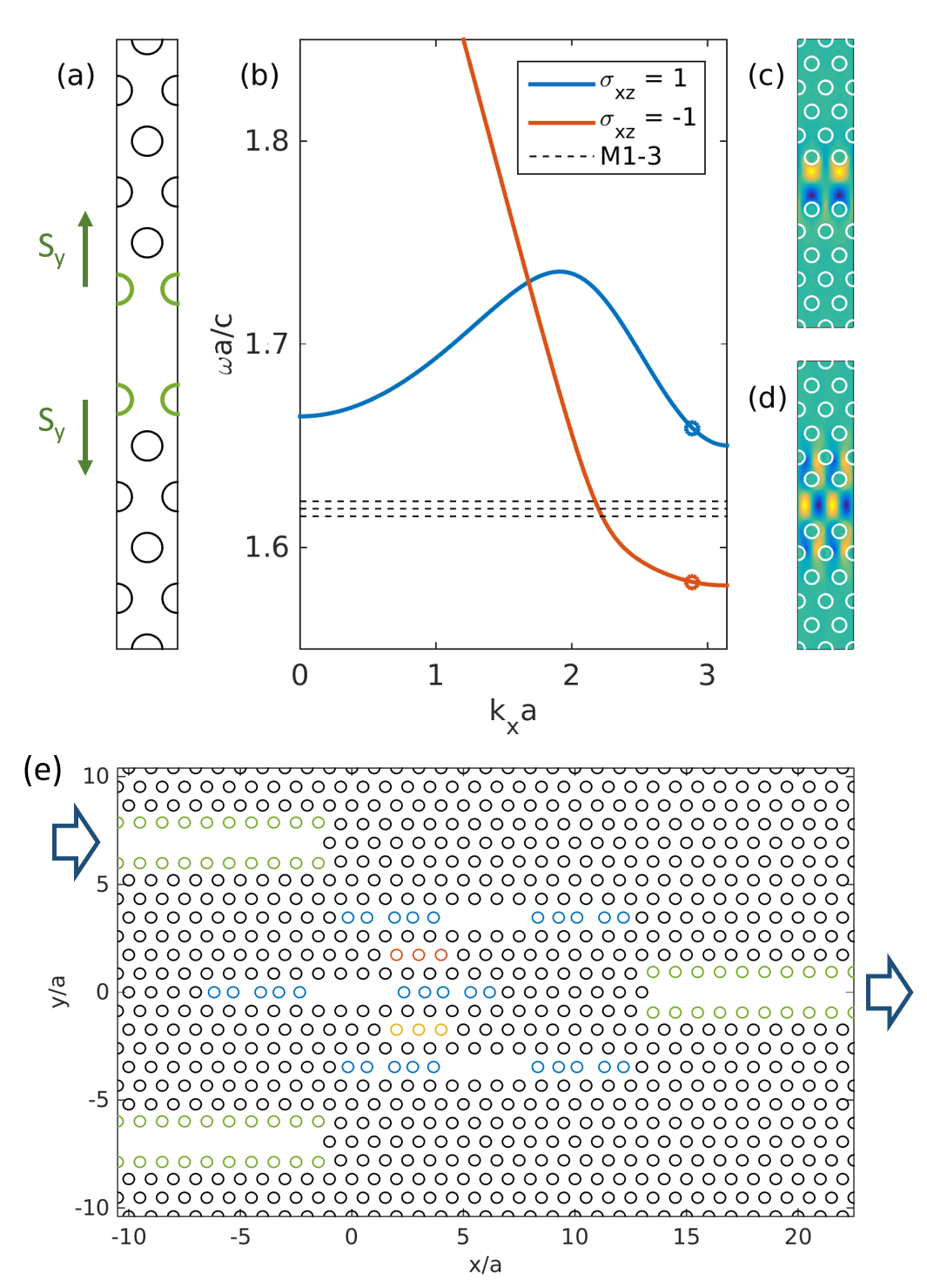}%
 \caption{(a): Unit cell of a PhC waveguide made of a missing row of holes. For dispersion tuning, the holes on each side of the missing row are shifted away by $S_{y} = 0.075a$. (b): Bloch bands of the waveguide; the modes of the blue (red) band are symmetric (anti-symmetric) with respect to the $xz$-plane bisecting the guide. The dashed lines show the frequencies of the three coupled-cavity modes. (c): Electric field $E_y$ profile of the mode at the frequency given by a blue circle in (b). (d): Same as (c), for the red circle in (b). (e): Input (left) and output (right) channels for the three-cavity system. The colored holes are shifted as defined in Fig. \ref{fig1}, and panel (a) here.}
\label{fig4}
\end{figure}

Next, we add an input and an output channel to the three-mode system. For this, we utilize the standard PhC waveguide that results from a missing row of holes in the crystal lattice (Fig. \ref{fig4}). The dispersion of the waveguide is plotted in Fig. \ref{fig4}(b). There are two bands, which are respectively symmetric and anti-symmetric with respect to the $\hat{\sigma}_{xz}$ reflection operator that was discussed earlier. The mode profile of one mode along the symmetric (blue) band is shown in panel (c), while that of an anti-symmetric mode is shown in panel (d). The guided bands can be tuned in frequency for example through changing the width of the guide by shifting the first row of holes away from the center (panel (a)). We design the device so that the three coupled-cavity modes (dashed lines in panel (b)) are resonant with the anti-symmetric band only, but not too close to the slow-light region at the band edge. This is achieved for a shift $S_y = 0.075a$, which was used to compute the dispersion of panel (b). In panel (e), we illustrate a possible integration of the cavities and the waveguides. The two guides on the left are used for pump input. They couple to all three modes, since the $xz$ planes bisecting each of the guides are not the same as the $xz$ plane bisecting the coupled-cavity system. We incorporate two guides into the design in order to preserve the structural symmetry with respect to the $y = 0$ plane. This plane is a symmetry of both the three-cavity system and the output waveguide on the right. Since the latter supports only anti-symmetric modes, only the signal and idler modes (M1 and M3) couple to this output channel, while the coupling of the pump mode (M2) is strictly zero. This comes from the symmetry consideration, but we will also verify it with a simulation in Section \ref{sec:disorder}.  

The FDTD-computed quality factors of the three-cavity system (without the waveguides) are $0.78 \times 10^6, 1.5 \times 10^6$, and $0.81 \times 10^6$, respectively. These were computed through the time-decay of the electric field in the center of $C1$, after selective excitation of each mode. Mode M2 can be easily singled out by imposing symmetric boundary conditions at the $y = 0$ plane. Modes M1 and M3 were selectively excited by imposing anti-symmetric boundary conditions, and by using narrow-band sources of pulse-length $1\mathrm{ps}$ centered around the two corresponding resonant frequencies. The eigenmodes of the coupled-cavity system thus have slightly lower $Q$'s than the $Q = 4.1 \times 10^6$ computed for a single optimized L3 cavity. In principle, a procedure similar to the one of Ref. [\onlinecite{Minkov2014}] can be used to re-optimize the $Q$-s of our system, but we consider the values sufficiently high as they are. Furthermore, we design the system to be in the over-coupled regime, in which the dominating losses are due to coupling to the input and output waveguides. This is an obvious requirement in terms of getting light in and out of the system, and, in addition, a controlled broadening of the resonant line-widths would help meeting the resonant condition experimentally. 

\begin{figure}
\centering
\includegraphics[width = 0.44\textwidth, trim = 0in 0in 0in 0in, clip = true]{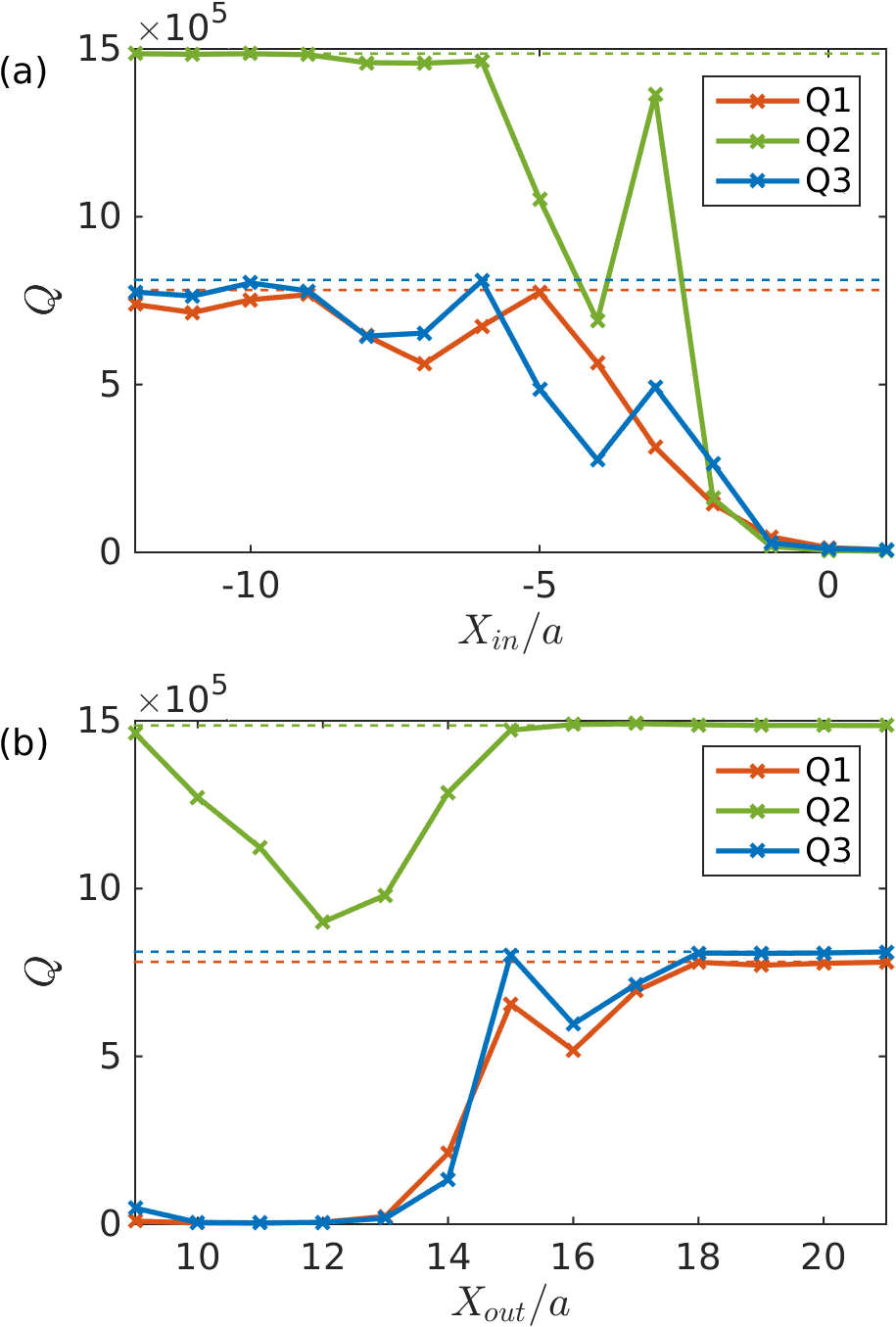}%
 \caption{(a): Quality factor of the three-cavity modes in the presence of input waveguides, versus the position $X_{in}$ of the latter. $X_{in}$ is defined as the position of the last missing hole, see Fig. \ref{fig4}(e). (b): Same as (a), but with the output waveguide.}
\label{fig5}
\end{figure}

\begin{figure*}[t]
\centering
\includegraphics[width = 1.0\textwidth, trim = 0in 0in 0in 0in, clip = true]{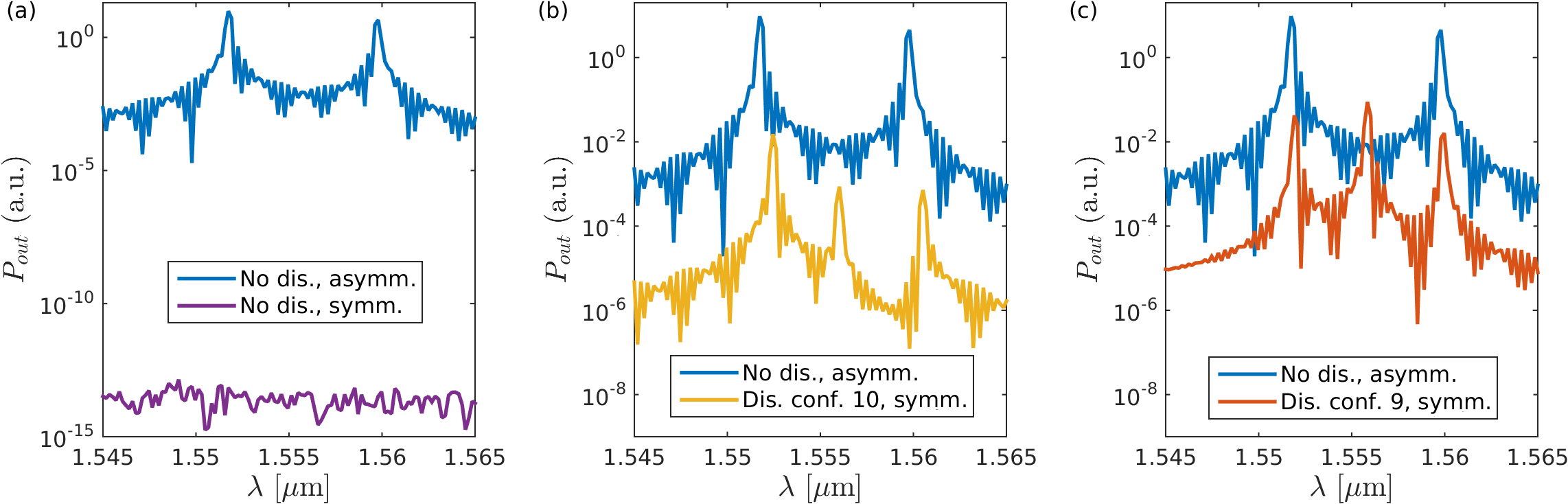}%
 \caption{Power spectra recorded in the output waveguide. Blue (in all panels): no disorder, asymmetric excitation. All the other data were recorded after symmetric excitation -- purple in (a): no disorder; yellow in (b): disorder configuration 10; red in (c): disorder configuration 9. Notice that the $y$-scale in (a) is different from that of (b) and (c).}
\label{fig6}
\end{figure*}

To study the coupling of the FWM modes into the waveguides of Fig. \ref{fig4}(e), we compute the $Q$-s exactly as for the three-cavity system, but this time in presence of either the input or the output waveguides. In Fig. \ref{fig5}(a), we show the dependence of the $Q$-s on the position of the input waveguides $X_{in}$, defined as the position of their last missing hole (e.g. $X_{in} = -2a$ in Fig. \ref{fig4}(e)). Both of the input waveguides are present in this simulation, while the output one is not. Conversely, in panel (b), we plot the $Q$-s in the presence of the output guide only, as a function of the position $X_{out}$ of its first missing hole ($X_{out} = 14a$ in Fig. \ref{fig4}(e)). As can be seen, the output waveguide affects $Q_2$ much more weakly than $Q_1$ and $Q_3$, due to the symmetry consideration. The fact that $Q_2$ changes at all is most likely due to extra out-of-plane scattering due to the perturbation introduced by the presence of the waveguide. For a good compromise between $Q$-s and waveguide coupling, in Section \ref{sec:disorder} below we fix the positions of the waveguides to $X_{in} = -2a$, $X_{out} = 14a$, i.e. the values illustrated in Fig. \ref{fig4}(e). This choice was made based on Fig. \ref{fig5}, where the two channels were analyzed separately. The quality factors of the three modes with both input and output waveguides were finally computed with FDTD, resulting in $Q_1 = 0.97 \times 10^5$, $Q_2 = 1.59 \times 10^5$, and $0.99 \times 10^5$. These values are perfectly matched by a model in witch the total losses are obtained by summing the intrinsic losses and the losses due to all the waveguides. In other words, if we define superscripts so that $Q^0$, $Q^{in}$, $Q^{out}$, and $Q^{tot}$ refer to the structure with no waveguides, input waveguides only, output waveguides only, or all waveguides, respectively, then $Q^{tot}$ for each of the three modes $i = 1, 2, 3$ can be computed as \cite{Minkov2013}

\begin{equation}
\frac{1}{Q^{tot}_i} = \frac{1}{Q^{in}_i} + \frac{1}{Q^{out}_i} - \frac{1}{Q^0_i}.
\end{equation}

The bare-cavity losses are subtracted in the end because they are actually counted twice when adding $Q^{in}$ and $Q^{out}$. The $Q^{tot}$ values obtained in this way are equal to the FDTD-computed values quoted above within the given precision, which illustrates that the input and output channels can be studied separately, as in Fig. \ref{fig5}. 
 
\section{Pump-filtering and disorder}
\label{sec:disorder}

To verify that there is indeed no power radiated into the output waveguide at the pump frequency, we use a power monitor in the FDTD simulation, placed at $x = 19a, y = 0$ (refer to Fig. \ref{fig4}(e)). In Fig. \ref{fig6}(a), we show the recorded power spectrum ($P_{out}$) upon excitation with two broadband sources centered at $C1$ and $C3$. The blue curve (which also appears in panels (b) and (c)) is recorded after antisymmetric excitation. The two peaks correspond to the signal and idler modes (i.e. the antisymmetric modes M1 and M3, see Fig. \ref{fig2})). The oscillations are due to the finite time duration of the simulation, which was again set to 25ps. To test the pump suppression, we further performed a simulation with a symmetric phase relation between the two sources. This configuration excites only $M2$, again due to symmetry. Consequently, the recorded $P_{out}$ shown in purple in Fig. \ref{fig6}(a) is at the negligible level below $10^{-13}$, and is likely due to numerical noise. This confirms the expected \textit{full filtering} of the pump mode in our proposed FWM design. 

In the presence of random structural disorder that breaks the $xz$-plane symmetry, a finite coupling of the pump mode to the output channel is expected. To quantify this effect, we perform simulations including random shifts in the position and radius of each hole, taken from a Gaussian distribution with zero mean and standard deviation $\sigma$. This commonly employed disorder model has been found to capture well the effects of fabrication imperfections \cite{Portalupi2011, Minkov2013}. We set $\sigma = 0.002a$, which is a typical magnitude for state-of-the-art PhCs\cite{Portalupi2011, Lai2014}, and simulate ten different disorder configurations. We use the symmetric excitation scheme, i.e. the one that results in \textit{no} output power in the disorder-less case. In the presence of disorder, however, an output power signal is recorded at all three mode frequencies. This is because the broken symmetry introduces both a finite excitation of the signal and idler modes, and a finite coupling of the pump mode to the output waveguide. Among the ten configurations, the two power spectra with the lowest and the highest $P_{out}$ at the pump frequency are shown in Fig. \ref{fig6}, panels (b) and (c), respectively. Notice that, while it is non-zero, there is still a significant suppression of 17 to 37 dB of the pump power when compared to the signal/idler output of the blue curve. 

\begin{figure*}[t]
\centering
\includegraphics[width = 1.0\textwidth, trim = 0in 0in 0in 0in, clip = true]{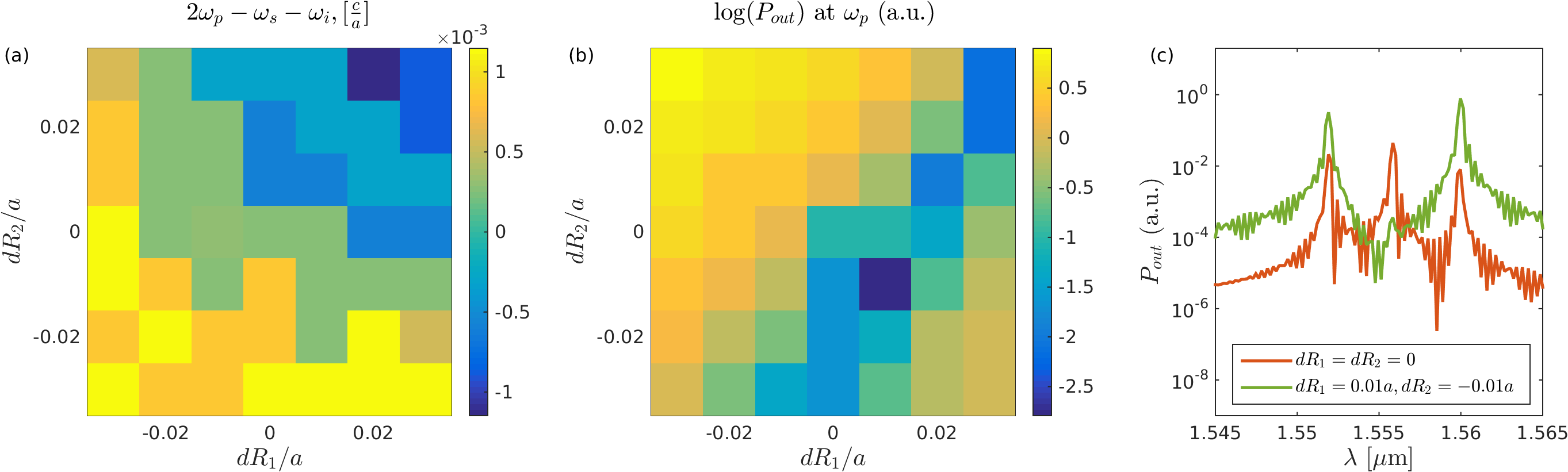}%
 \caption{Tuning of the disordered system (configuration 9) using the parameters $dR_1$ and $dR_2$. (a): The frequency relation $2\omega_p - \omega_s - \omega_i$, which has to be close to zero for FWM. (b): Pump power radiated into the output waveguide. (c): Power spectrum in the output waveguide for two different $dR_1$, $dR_2$ values. The red plot is the same as the one in Fig. \ref{fig6}(c). The green curve represents the `fixed' filtering, with $dR_1$ and $dR_2$ taken from panel (b) such that the pump power is the lowest.}
\label{fig7}
\end{figure*}

We now explore the possibility to tune the device through the shifts $dR_1$ and $dR_2$, in order to counter some of the disorder effects. We take disorder configuration 9 (red curve in Fig. \ref{fig6}), as it has the strongest power at the pump frequency. Notice that we must pay attention to keep the frequencies within the resonant condition. Because of this, in Fig. \ref{fig7}(a) we plot the quantity $2\omega_p - \omega_i - \omega_s$ as a function of $dR_1$ and $dR_2$. This quantity is still close to zero (equal within the simulation error due to the finite time) for the nominal case of $dR_1 = dR_2 = 0$. In other words, disorder, as well as the presence of the waveguides, did not interfere strongly with the resonance condition. Notice however that there is a whole range of $dR_1$, $dR_2$ values for which the condition is also met. Thus, in panel (b) of Fig. \ref{fig7}, we plot $P_{out}$ at the pump frequency as a function of these two shifts. This varies over three orders of magnitude in the plotted range, illustrating the potential for improvement of the filtering. The minimum value is achieved for $dR_1 = 0.01a$, $dR_2 = -0.01a$, for which in fact the resonant condition is still met (see panel (a)). Thus, a slight tuning of the $C_1$-$C_2$ and the $C_2$-$C_3$ couplings can lead to a significant improvement in filtering. This is also illustrated in panel (c), where we plot two power spectra as in Fig. \ref{fig6} -- without and with the tuning. The difference at the pump wavelength between the two curves is 21 dB, meaning that a total pump suppression of 38 dB with respect to the signal and idler output (blue curve in Fig. \ref{fig6}) can be achieved. 

\section{Discussion}

A remark on the opposite signs of $dR_1$ and $dR_2$ for the best pump suppression is due. This result makes sense, since coupling to the output channel is not introduced by disorder per se, but by the breaking of the $\hat{\sigma}_{xz}$ symmetry. In other words, if we would introduce disorder invariant under that reflection, the full filtering would be preserved. Realistic disorder, however, does not have that property, and the purpose of tuning $dR_1$ and $dR_2$, to a first approximation, is to restore that symmetry as much as possible. Thus, asymmetric values for the two quantities are logical. In practice, depending on the tuning method, it might be that either only positive or only negative values are achievable. In this case, one should ideally start from a slightly smaller (respectively larger) nominal hole radius of the six holes involved, in order to fully benefit from the parameter space spanned by $dR_1$ and $dR_2$.

The efficiency of a resonant non-linear processes like the FWM discussed here generally increases with an increase in the concentrated electromagnetic energy\cite{Barclay2005, Lin2007}. This increases with the quality factor of the modes, and, loosely speaking, decreases with the mode volume. The latter has no unique definition, but instead an effective quantity can be defined for various applications. More specifically, the FWM efficiency is inversely proportional to an integral that takes into account the overlap between all three modes\cite{Lin2007, Zeng2014}. Qualitatively, our system is similar to the one experimentally characterized in Ref. [\onlinecite{Azzini2013}], where three coupled PhC nanobeam cavities were used. In that work, the measured spontaneous generation rate was $\approx 300$(MHz/mW$^2$)$P_p^2$, with $P_p$ the pump power expressed in mW. This is almost two orders of magnitude higher than that of micro-ring systems: for example, this same quantity is quoted as 5(MHz/mW$^2$)$P_p^2$ in Ref. [\onlinecite{Azzini2012}]. The generation rate is not explicitly quoted in Refs. [\onlinecite{Clemmen2009},\onlinecite{Harris2014}], but can be inferred from the data presented there, and has a similar value of $\approx 7$(MHz/mW$^2$)$P_p^2$. The L3 cavities used here have a mode volume similar to that of the nanobeam cavities of Ref. [\onlinecite{Azzini2013}] (compare Refs. [\onlinecite{Minkov2014}] and [\onlinecite{Velha2007}]). Thus, the generation rate is expected to be similar, given comparable quality factors. In other words, a strong improvement when compared to micro-ring systems can reasonably be expected in our device simply due to the stronger light confinement. Beyond that, however, the quality factors of the modes of our system are more than an order of magnitude higher than those of Ref. [\onlinecite{Azzini2013}], and the FWM efficiency is proportional to $Q^3$ for modes of approximately equal $Q$\cite{Lin2006, Azzini2012}. In short, our expected efficiency is three orders of magnitude higher than that of Ref. [\onlinecite{Azzini2013}], which is already close to two orders of magnitude higher than in micro-rings. 

This consideration is qualitative, but it makes it very reasonable to expect the efficiency of our system to be \textit{at the very least} two orders of magnitude larger than that of Ref. [\onlinecite{Harris2014}], the only work where on-chip filtering was achieved. For this purpose, a long DBR was used, providing a pump suppression of 65 dB while only weakly affecting the signal and idler modes. In contrast, our device is fully self-filtering in the absence of disorder, i.e. the pump transmission is strictly zero due to symmetry. Beyond that, we have shown that even in presence of disorder, a pump suppression of close to 40 dB can be achieved in state-of-the art silicon systems, especially if some tuning of the cavity-cavity coupling is possible. This, together with the higher non-linear efficiency that can be expected from our device, means that for the same generation rates as in Ref. [\onlinecite{Harris2014}], the DBR would most probably be unnecessary, and thus the footprint of an integrated setup would be significantly reduced. Additionally, the signal and idler losses involved in guiding the light through the DBR or any other filtering device would also be eliminated. Furthermore, the idea of symmetry-imposed filtering could stimulate novel proposals in which the signal and idler \textit{themselves} can be demultiplexed without the need for add-drop filters. This would again decrease both the footprint and the losses of an integrated device. Finally, apart from the pump-filtering advantage, our design is also expected to have a higher photon generation efficiency, which is anyway beneficial as it leads to a lower power consumption and/or higher brightness of the source. 

In summary, we have presented a photonic crystal coupled-cavity system optimized for photon-pair generation through a resonant FWM process. The main innovation is the self-filtering of the pump, which does not couple to the output channel due to symmetry. The filtering of the pump power is one of the main challenges for the full on-chip integration of FWM-based photon sources. Thus, our device, which also promises a higher conversion efficiency when compared to microring resonators, is an attractive component for future integrated quantum technologies.  

This work was supported by the Swiss National Science Foundation through Project N\textsuperscript{\underline{o}} 200020\_149537.

\bibliography{fwm}

\end{document}